\documentclass[aps,prl,twocolumn,showkeys,superscriptaddress]{revtex4-1}

\usepackage[artemisia]{textgreek}
\usepackage{epsfig}
\usepackage{amsmath}
\usepackage{graphicx}
\usepackage{dcolumn}
\usepackage{latexsym}
\usepackage{color}
\usepackage{epstopdf}
\usepackage{subfigure}
\usepackage{comment}
\usepackage{nicefrac}
\usepackage{blindtext}
\usepackage[ulem=normalem]{changes}
\usepackage{float}
\usepackage{xcolor}
\usepackage{soul}

\newcommand*{\citen}[1]{%
  \begingroup
    \romannumeral-`\x 
    \setcitestyle{numbers}%
    \cite{#1}%
  \endgroup   
}

\begin{document}
\title{Engineering Hybrid Epitaxial InAsSb/Al Nanowire Materials for Stronger Topological Protection}\textbf{}
\author{Joachim E. Sestoft\textsuperscript{\textdagger}}
\affiliation{Center for Quantum Devices and Station Q Copenhagen, Niels Bohr Institute, University of Copenhagen, 2100 Copenhagen, Denmark}
\altaffiliation{These authors contributed equally to this work.}

\author{Thomas Kanne\textsuperscript{\textdagger}}
\affiliation{Center for Quantum Devices and Station Q Copenhagen, Niels Bohr Institute, University of Copenhagen, 2100 Copenhagen, Denmark}
\altaffiliation{These authors contributed equally to this work.}

\author{Aske N\o rskov Gejl\textsuperscript{\textdagger}}
\affiliation{Center for Quantum Devices and Station Q Copenhagen, Niels Bohr Institute, University of Copenhagen, 2100 Copenhagen, Denmark}
\altaffiliation{These authors contributed equally to this work.}

\author{\mbox{Merlin von Soosten}}
\affiliation{Center for Quantum Devices, Niels Bohr Institute, University of Copenhagen, 2100 Copenhagen, Denmark}

\author{\mbox{Jeremy S. Yodh}}
\affiliation{Center for Quantum Devices and Station Q Copenhagen, Niels Bohr Institute, University of Copenhagen, 2100 Copenhagen, Denmark}

\author{Daniel Sherman}
\affiliation{Center for Quantum Devices and Station Q Copenhagen, Niels Bohr Institute, University of Copenhagen, 2100 Copenhagen, Denmark}

\author{Brian Tarasinski}
\affiliation{QuTech and Kavli Institute of Nanoscience, Delft University of Technology, 2600 GA Delft, The Netherlands}

\author{Michael Wimmer}
\affiliation{QuTech and Kavli Institute of Nanoscience, Delft University of Technology, 2600 GA Delft, The Netherlands}

\author{\mbox{Erik Johnson}}
\affiliation{Center for Quantum Devices, Niels Bohr Institute, University of Copenhagen, 2100 Copenhagen, Denmark}
\affiliation{Department of Wind Energy, Technical University of Denmark, Ris\o{} Campus, 4000 Roskilde, Denmark}

\author{\mbox{Mingtang Deng}}
\affiliation{Center for Quantum Devices and Station Q Copenhagen, Niels Bohr Institute, University of Copenhagen, 2100 Copenhagen, Denmark}

\author{Jesper Nyg\aa rd}
\affiliation{Center for Quantum Devices, Niels Bohr Institute, University of Copenhagen, 2100 Copenhagen, Denmark}

\author{Thomas Sand Jespersen}
\affiliation{Center for Quantum Devices, Niels Bohr Institute, University of Copenhagen, 2100 Copenhagen, Denmark}

\author{Charles M. Marcus}
\affiliation{Center for Quantum Devices and Station Q Copenhagen, Niels Bohr Institute, University of Copenhagen, 2100 Copenhagen, Denmark}

\author{Peter Krogstrup}
\email{krogstrup@nbi.dk}
\affiliation{Center for Quantum Devices and Station Q Copenhagen, Niels Bohr Institute, University of Copenhagen, 2100 Copenhagen, Denmark}

\date{\today}
\begin{abstract}

The combination of strong spin-orbit coupling, large $g$-factors, and the coupling to a superconductor can be used to create a topologically protected state in a semiconductor nanowire. Here we report on growth and characterization of hybrid epitaxial InAsSb/Al nanowires, with varying composition and crystal structure. We find the strongest spin-orbit interaction at intermediate compositions in zincblende InAs$_{1-x}$Sb$_{x}$ nanowires, exceeding that of both InAs and InSb materials, confirming recent theoretical studies \cite{winkler2016topological}. We show that the epitaxial InAsSb/Al interfaces allows for a hard induced superconducting gap and 2$e$ transport in Coulomb charging experiments, similar to experiments on InAs/Al and InSb/Al materials, and find measurements consistent with topological phase transitions at low magnetic fields due to large effective $g$-factors. Finally we present a method to grow pure wurtzite InAsSb nanowires which are predicted to exhibit even stronger spin-orbit coupling than the zincblende structure.


\end{abstract}
\pacs{}
\keywords{InAsSb, nanowires, hybrid epitaxial materials, spin-orbit coupling, topological superconductivity, semiconductor-superconductor epitaxy, majorana zero modes}
\maketitle

Semiconductor-superconductor materials hold potential for a variety of gateable superconducting experiments and applications. Semiconductor nanowires (NWs) with strong spin-orbit interaction (SOI) coupled to a superconductor are of particular interest \cite{oreg2010helical,lutchyn2010majorana,sarma2015majorana} due to the prospect of hosting topologically protected Majorana zero modes (MZM) which can be used for fault tolerant quantum computing \cite{freedman2003topological,read2000paired,kitaev2001unpaired,fu2008superconducting}.~So far, signatures of MZM have been reported on binary III-V semiconductor/superconductor hybrids,~e.g.~InSb/Nb(TiN) and InAs/Al NWs \cite{mourik2012signatures,deng2012anomalous,das2012zero,deng2016majorana,albrecht2016exponential}, which possess the necessary properties for realizing a topological superconductor. Besides induced superconductivity, realizing a topological protection and MZMs requires an applied magnetic field with a magnitude that depends on the effective SOI and $g$-factor of the hybrid system. Thus, realizing strong topological protection depends on the ability to engineer materials with the appropriate properties. A material platform which remains unexplored is ternary based semiconductors such as InAsSb \cite{namazi2017direct}, which have been predicted to exhibit much stronger spin-orbit coupling than its binary compounds \cite{winkler2016topological}, and could potentially provide a material with sufficiently strong topological protection to realize topological quantum information applications.


In this letter, we present structural, compositional and electronic characterization of zincblende (ZB) and wurtzite (WZ) InAs$_{1-x}$Sb$_{x}$ NWs with and without epitaxially grown Al.~We characterize the crystal structure and composition with high-resolution transmission electron microscopy (TEM) methods, and the electronic properties with low temperature transport measurements. From measurements of weak anti-localization (WAL) on segmented InAs$_{1-x}$Sb$_{x}$ NWs of different Sb compositions $x$, we find a non-monotonic dependence of spin-orbit length ($l$\textsubscript{SO}) on the composition, with the smallest measured $l$\textsubscript{SO} at $x$ $\sim$ 0.5. For the hybrid InAsSb/Al NWs we find characteristic epitaxial semiconductor-superconductor structural ordering, where 3:2 lattice matched interfacial domains appear as the preferred interface for all compositions characterized. However, the interfaces do not appear as atomically sharp as in the case of the epitaxial InAs/Al interfaces \cite{krogstrup2015epitaxy}, and residual dislocation arrays are present in the semiconductor as a result of strain relaxation relative to the strongly bound bi-crystal interface. Despite these structural details, we do not observe any degradation of the induced gap from tunnel spectroscopy measurements on InAs\textsubscript{0.2}Sb\textsubscript{0.8}/Al NWs.~We study the density of states (DOS) as a function of applied magnetic field and show that Andreev bound states (ABS) merge to zero energy at relatively low magnetic fields due to large effective $g$-factors of the hybrid system. This is consistent with a topological phase transition as expected for materials with large SOI \cite{sarma2015majorana,deng2016majorana,lutchyn2017realizing}. Coulomb charging experiments performed on NWs from the same growth batch show a transition from 2$e$ to 1$e$ charge periodicity at magnetic fields comparable to where the zero energy states are observed in the DOS measurements. InAsSb NWs with ZB structure at intermediate concentrations of Sb were not measured as it was difficult to gate and pinch off with standard gating geometries. To circumvent this we present a method to grow WZ InAs\textsubscript{0.7}Sb\textsubscript{0.3}/Al NWs, which is relevant both because of a lower electron affinity compared to ZB, and because of a potentially higher SOI. We find an improved electrostatic gate response and measure a hard induced superconducting gap, comparable to what is measured in ZB NWs.

\section{Results and Discussion}

\begin{figure*}[ht!]
\vspace{0.2cm}
\includegraphics[scale=1]{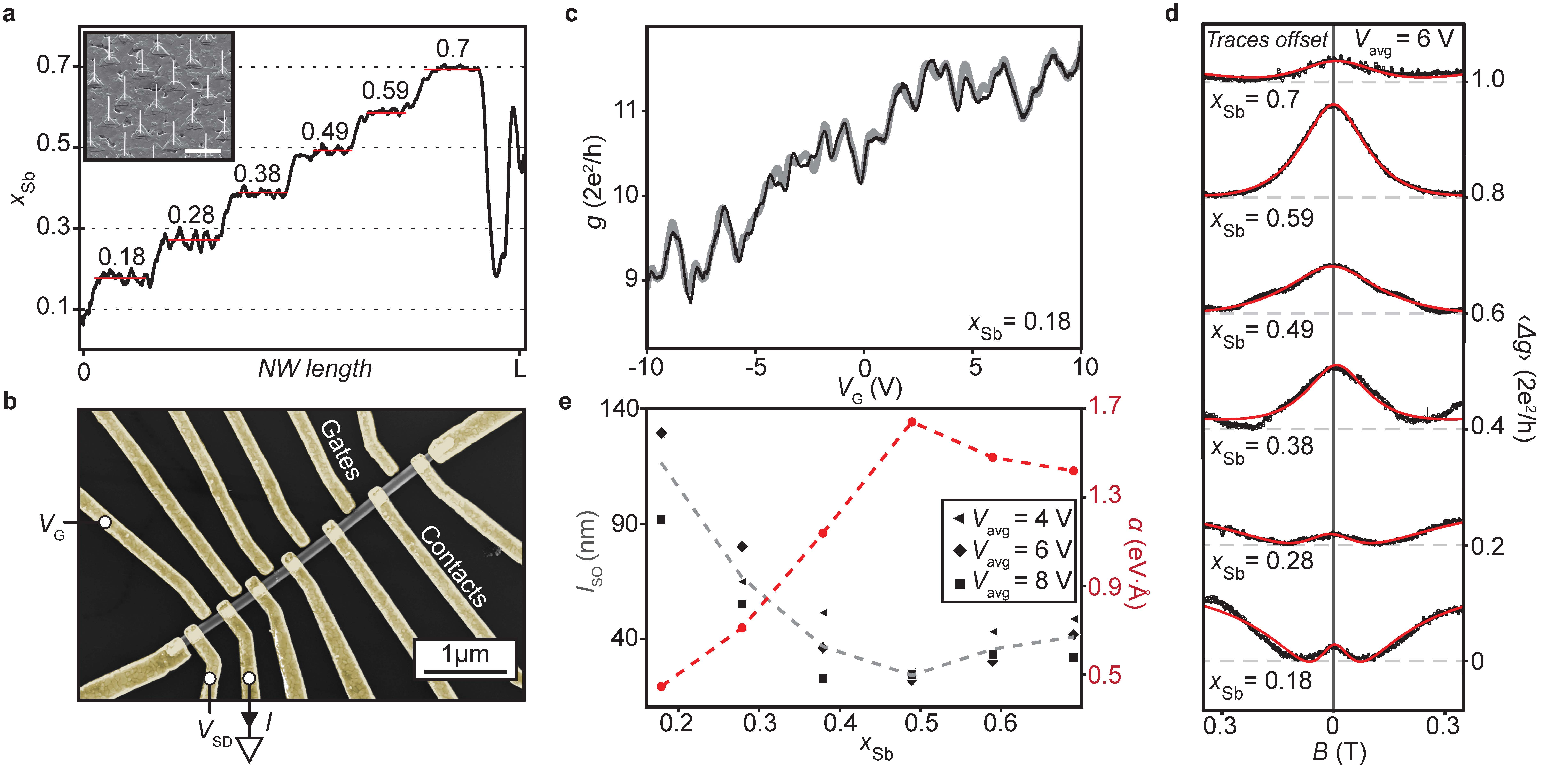}
\vspace{0.2cm}
\caption{\textbf{Spin-orbit interaction in InAs$_{1-x}$Sb$_{x}$ NWs.} \textbf{a}, Sb concentration, $x$, as a function of NW length, $L$, with a step-like profile of the As/Sb molar fraction along the NW. The inset shows the nanowires as grown on the substrate. \textbf{b}, False-colored scanning electron micrograph of a typical device. Yellow, Ti/Au contacts and gates; grey, InAs$_{1-x}$Sb$_{x}$ NW; $V$\textsubscript{SD}, the applied voltage bias; $I$, measured current; $V$\textsubscript{G}, gate voltage controlling the chemical potential on individual segments. \textbf{c}, Trace/retrace (black/grey) of differential conductance for $x$ = 0.18 as a function of $V$\textsubscript{G} showing aperiodic reproducible fluctuation with amplitude $\sim$ e\textsuperscript{2}/h. \textbf{d}, Averaged differential conductance change, $\langle \Delta g \rangle$ as a function of magnetic field showing suppression of the weak anti-localization effect around $B$ = 0 for all concentrations. Red overlay is the fit from eq. 1, from which $l$\textsubscript{SO} is extracted. \textbf{e}, Spin-orbit length (left y-axis) as a function of Sb concentration. A noticeable minimum in $l$\textsubscript{SO} is observed around $x$ $\sim$ 0.5, for the three different gate averaging amplitudes, $V$\textsubscript{avg} = 4 V, 6 V and 8 V. The Rashba coefficient (right y-axis) is a conversion of the average $l$\textsubscript{SO} displaying a qualitative maximum around $x$ $\sim$ 0.5. 
}
\label{fig1}
\end{figure*}

To study the dependence of SOI strength on composition, we measure the magnetoconductance in axially segmented heterostructure NWs with six different compositions along the NW length. The NWs have a step-like compositional change along the growth direction as seen on the energy dispersive x-ray spectroscopy (EDX) line-scan in Fig. 1 \textbf{a}. The quantification of composition with TEM-EDX is done following the approach presented in the supplementary information of Ref. \cite{Krogstrup2009}, and was confirmed with relative lattice spacing measures from high resolution TEM (assuming Vegard's law). From measurements on multiple NWs, we find that each segment has a given length relative to the total NW length, which makes it possible to contact each individual segment by electron beam lithography (EBL), as seen in Fig. 1 \textbf{b}. Each contact is placed on the transition from one composition to the next, ensuring that every segment of the device has a given composition, see supplementary information S3 for details on the device fabrication. All measurements were carried out using standard a.c. lock-in measurements in a dilution refrigerator with a base temperature of $\sim$ 35 mK. 

Two-terminal differential conductance measurements as a function of gate voltage, $V$\textsubscript{G}, were performed on each segment. An example of a trace/retrace is displayed for $x$ = 0.18 in Fig.~1\textbf{c}. Reproducible conductance fluctuations are observed with an amplitude on the order of e\textsuperscript{2}/h indicating universal conductance fluctuations (UCF). This suggests that the phase coherence length, l\textsubscript{$\phi$}, is on the order of or longer than the device length. We assume that the elastic scattering length, l$_e$, is shorter or comparable to l\textsubscript{$\phi$} which means that the transport resides in the diffusive regime \cite{altshuler1985fluctuations,hansen2005spin,PhysRevLett.55.1622, lee1987universal}. A measure of the spin-orbit length is then extracted by fitting the magnetoconductance measurements, as similarly done on InAs and InSb NWs \cite{hansen2005spin, hernandez2010spin, roulleau2010suppression, dhara2009magnetotransport, van2015spin, wang2015phase}. 
In this work the contribution from UCF was averaged out by modulating the side gates with amplitudes, V\textsubscript{avg}, using a sawtooth shaped wave with a 2 Hz frequency. Here the averaging of the differential conductance was measured with a 372 Hz lock-in modulation over a 3 s period \cite{sand2017}. The typical scale of the UCF was found from the correlation function to be $V$\textsubscript{c} $\sim$ 0.5 V, and by using $V$\textsubscript{avg} $\gg$ $V$\textsubscript{c} we ensure effective averaging over many UCF periods. 

The average differential conductance, $\langle \Delta g\rangle $, is plotted as a function of magnetic field in Fig. 1 \textbf{d} for all device segments using $V$\textsubscript{avg} = 6 V. A characteristic increase in $\langle \Delta g \rangle$ symmetrically around $B$ = 0 is observed indicating WAL which is canceled away from zero field \cite{hikami1980spin, bergman1982influence}. In order to extract a measure of the phase coherence length, $l$\textsubscript{$\phi$}, and
the spin-orbit scattering length, $l$\textsubscript{SO}, we fit the measured magnetoconductance traces with a commonly used expression for the WAL correction to the conductance. We note that the extracted length scales are dependent on the model and fitting approach, rather than quantitative measures. However, we can reasonably compare relative length scales between the different compositions, and compare to literature values that uses the same fitting approach. The model used here for the correction in the diffusive limit for small elastic scattering lengths is expressed as \cite{kurdak1992quantum, al1981magnetoresistance}, 

\begin{equation}\label{eq1}
\begin{split}
\Delta g(B) = -\frac{2e^2}{hL} \biggl[\frac{3}{2}\left(\frac{1}{l_{\phi}^2} + \frac{4}{3l_{\text{SO}}^2} + \frac{1}{D\tau_{\text{B}}} \right)^{-1/2} \\ 
-\frac{1}{2}\left(\frac{1}{l_{\phi}^2}+\frac{1}{D\tau_\text{B}}\right)^{-1/2}\biggr]. 
\end{split}
\end{equation}

\noindent Here, the magnetic dephasing time ($\tau$\textsubscript{B}) is defined as $\tau_{\text{B}} = \frac{C l_{\text{m}}^4}{W^2D}$ where D is the diffusion constant and the magnetic length, $l$\textsubscript{m} $= \sqrt{\frac{\hbar}{eB}}$, is on the order of the NW diameter for the fitting range 0.35 T, as in Ref. \citen{hansen2005spin, roulleau2010suppression, dhara2009magnetotransport}. 

The prefactor, $C$, can in principle be computed numerically for a given geometry\cite{van2015spin}. In general, it depends on details of the systems such as whether there is a surface accumulation. However, changes in the geometry change $C$ typically only by a factor of order 1, as we show on an example in the supplementary information S8. Since we are mainly interested in the qualitative behavior of the SOI, we follow previously published work, and use $C$ = 3 \cite{beenakker1991quantum, hansen2005spin}. Taking a fixed nanowire diameter $W$ = 100 nm (as determined by SEM), we then use Eq. \ref{eq1} to fit the magnetoconductance data measured with averaging voltages $V$\textsubscript{avg} = 4, 6 and 8 V. The red overlay in Fig. 1 \textbf{d} shows the fit to the WAL data for all segments at $V$\textsubscript{avg} = 6 V. We note that $l$\textsubscript{$\phi$} shows no apparent dependence on composition (see Supplementary Information S4 for details). The extracted $l$\textsubscript{SO} is shown in Fig.~1 \textbf{e} as a function of composition for all applied averaging voltages. Here, the smallest $l$\textsubscript{SO} is obtained at $x$ $\sim$ 0.5, indicating a non-monotonic dependence on composition and that a significantly smaller $l$\textsubscript{SO} can be achieved compared to pure InAs and InSb NWs \cite{hernandez2010spin,van2015spin,hansen2005spin,dhara2009magnetotransport}.

As these NWs are grown along the [111]-direction with a pure ZB crystal structure, the Dresselhaus contribution from bulk inversion asymmetry is negligible \cite{van2015spin} and therefore dominated by the Rashba contribution from structural inversion asymmetry. Therefore, we relate $l$\textsubscript{SO} to the Rashba coefficient $\alpha$, which is a measure of the Rashba spin-orbit strength. Following Ref. \citen{roulleau2010suppression, schapers2006suppression}, we can rewrite the relation between the $l$\textsubscript{SO} in a diffusive system into the spin-precession length $l$\textsubscript{R} which is related to the Rashba coefficient as $l_R = \frac{\hbar^2}{2m^{\star}\alpha_R}$. Using the estimate of the relation between $l$\textsubscript{SO} = $\frac{\sqrt[]{3}l_R^2}{W}$ we can relate the Rashba coefficient to the spin-orbit scattering lengths as, $\alpha_R = \frac{\hbar^2}{2} m^{\star -1}(\frac{l_{SO}W}{\sqrt[]{C}})^{-1/2}$. Estimates of $m^\star$ as a function of composition are found in Ref. \cite{vurgaftman2001band}. In Fig. 1 \textbf{e} we plot $\alpha$ as a function of composition and find the highest measured value of 1.65 eV$\cdot$\AA{} at $x$ $\sim$ 0.5. This approach must be applied with caution since the obtained $l$\textsubscript{R} is smaller than the diameter of the NW. However, for qualitative analysis of the Rashba contribution we expect this model to be applicable. 

The strong SOI at intermediate compositions can have different origins. One explanation is that a small band gap enhances the Rashba SOI, because the asymmetry from the valence band has a stronger hybridization through virtual processes between the conductance and valence bands \cite{winkler2003origin}. Thus, one could expect a maximum SOI around $x$ $\sim$ 0.6, where InAsSb has a minimum band gap \cite{yeh1992zinc,winkler2003origin}. However, ordering effects could also lead to SOI enhancement effects, as shown in recent ab-initio modeling performed on InAs$_{1-x}$Sb$_{x}$ NWs \cite{winkler2016topological}. Here a strongly enhanced SOI is found in the case of CuPt ordered stacking for compositions around $x$ = 0.5. We did not find any signature of such ordering in our TEM characterization, but small tendencies for such ordering could be reflected in the measurements of the SOI, switching the minimum $l$\textsubscript{SO} towards $x$ = 0.5. Finally, it should be noted that previous studies have shown that the SOI measured by WAL can be influenced by the carrier density. Due to a weak gate-coupling in these devices we cannot separate the carrier density and mobility of the samples. Thus the variation of $l$\textsubscript{SO} may also be influenced by a dependence of carrier density on $x$.

\begin{figure}[ht!]
\vspace{0.2cm}
\includegraphics[scale=1]{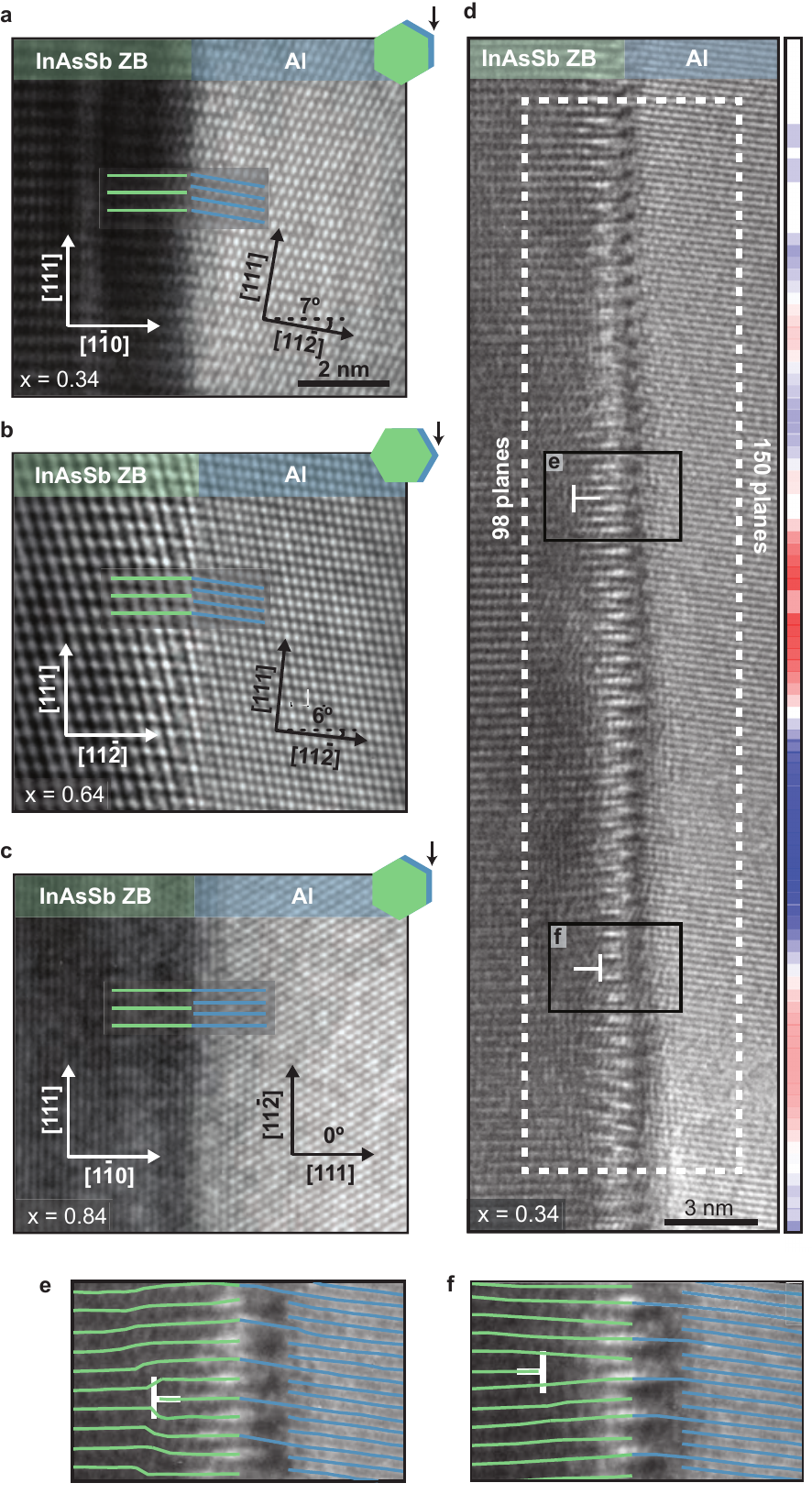}
\vspace{0.2cm}
\caption{\textbf{Bi-crystal interfacial matches of InAs$_{1-x}$Sb$_{x}$/Al NWs.}~\textbf{a}, 7$^{\circ}$ rotation between the [1$\bar{1}$0] and the [11$\bar{2}$]-direction in the semiconductor and the superconductor at $x$ = 0.34, respectively. Scale bar is the same for \textbf{a}-\textbf{c}. \textbf{b}, Rotation between the two [11$\bar{2}$]-directions for $x$ = 0.64 is 6$^{\circ}$.~\textbf{c}, No observed rotation between the [1$\bar{1}$0] and [111] for $x$ = 0.84.~\textbf{d}, Bi-crystal Burger circuit on a long segment of an interface for $x$ = 0.34 showing two edge dislocations with respect to 3:2 domain match. Color bar shows relative bending of the inter-plane distances with respect to a reference region away from the interface. Blue is upwards bending, red is downwards bending.~\textbf{e, f}, The two dislocations from \textbf{d}.}
\label{fig2}
\end{figure}

In the following section we focus on epitaxial growth of Al on InAsSb NWs at different compositions.~As previously shown, a thin Al shell can be grown epitaxially on selected facets of InAs NWs at low temperatures \cite{krogstrup2015epitaxy, gusken2017mbe, gazibegovic2017epitaxy}, with a resulting hard induced superconducting gap \cite{chang2015hard}, which makes them promising materials for gatable superconducting devices with low quasi-particle poisoning rates \cite{higginbotham2015parity}. However, correlations between structural properties of the epitaxial match and the associated induced superconducting properties have so far not been studied in detail. By introducing Sb we change the lattice constant and crystal structure, which affects the relative orientation of the Al and bi-crystal interfacial match. For NWs grown in the conventional [111]B/[0001]B growth direction, the crystal structure changes from WZ to ZB, when exceeding a given small fraction of Sb \cite{potts2015twinning}. Also, under the growth conditions used here, the facets change from InAs $\{1\bar{1}00 \}$\textsubscript{WZ} to InAsSb $\{1\bar{1}0 \}$\textsubscript{ZB}, i.e. rotating 30$^{\circ}$, and the lattice spacing increases linearly as a function of Sb concentration \cite{lugani2012modeling, ercolani2012growth}. 

In Fig.~\ref{fig2} \textbf{a}-\textbf{c} we show different bi-crystal interfaces for different InAsSb compositions. The white and black arrows specify the crystal orientations for the InAsSb and Al, respectively. The schematics in the top-right corners indicate the viewing direction in relation to the NW morphology. Characteristic orientations of the Al grains are observed to be strongly dependent on the InAsSb composition. For low Sb concentrations, $x \sim 0.34$ (Fig. \ref{fig2}, \textbf{a}) we find in-plane interfacial domains, which can be expressed in compact notation \cite{krogstrup2015epitaxy}, $\label{eq_0_34}
\Big( \frac{3_{Al,[ 111]}}{2_{InAsSb, [ 111] }} , -2.03\% \Big) \| \times \Big( \frac{8_{Al,[ 1 \bar{1} 0]}}{9_{InAsSb, [ 11 \bar{2}] }}, 0.55\% \Big) \bot,
$ where the transverse domain match is estimated from atomic lattice modelling (see Supporting Information S4). As the Sb concentration is increased (Fig. \ref{fig2} \textbf{b}) we find that the Al phase keeps an orientation that preserves an interfacial match with a 3:2 axial domain, $ \label{eq_0_64} 
\Big( \frac{3_{Al,[ 111]}}{2_{InAsSb, [ 111] }} , -3.99\% \Big) \| \times \Big( \frac{8_{Al,[ 11 \bar{2} ]}}{5_{InAsSb, [ 11 \bar{2}] }}, 2.41\% \Big) \bot.$ However, the transverse match changes by aligning the Al $[ 11\bar{2}]$ planes along the semiconductor $[ 11\bar{2}]$ planes. By increasing the Sb concentration even further (Fig. \ref{fig2} \textbf{c}) the Al phase attains the low interface energy domain of 3:2, with a low residual axial mismatch, $ \label{eq_0_84} 
\Big( \frac{3_{Al,[ 11\bar{2}]}}{2_{InAsSb, [ 111] }} , 0.5\% \Big) \| \times \Big( \frac{8_{Al,[ 1 \bar{1} 0]}}{9_{InAsSb, [11 \bar{2}] }}, -2.75\% \Big) \bot.
$

The Al rotates discretely depending on the Sb concentration, with rotations occurring both radially and axially around the ($11\bar{2}$) or ($111$) rotational axes. Examples of radial rotations are highlighted in Fig. \ref{fig2} \textbf{a} and \textbf{b}, whereas \ref{fig2} \textbf{c} shows axial rotation but no radial rotation. Even small axial rotations obscure the visualization of the bi-crystal match when imaging with TEM, because the InAsSb and Al will not simultaneously have a high symmetry zone-axis. The Al orientation can be understood in terms of minimization of thermodynamic excess free energy under the constraints of kinetic barriers during grain growth \cite{krogstrup2015epitaxy}. The four main terms that contribute to the excess of the chemical potential is the surface energy, the grain boundaries, the semiconductor/superconductor interface and related strain energies. Rotations increase the relative lattice plane distances at the interface and therefore may decrease the residual interfacial mismatch and lead to a higher ordered epitaxial interface and lower semiconductor/superconductor interfacial energy.~If the interfacial bonding is strong, the semiconductor/superconductor interface term may dominate and lead to rotations in order to minimize the interfacial bonding energy. We propose that the discrete characteristics of the rotations are due to the fact that specific rotations gives high symmetry cutting planes with low energy interfaces (see supplementary information S5). 

As shown above we generally find that the axial interfacial domain match seeks a 3:2 relation even though other lower ordered domains would give a lower residual mismatch. The 3:2 domain match is seen in the periodic interference effect (light/dark repeating pattern) along the interface. The dashed white rectangle in Fig. \ref{fig2} \textbf{d}, shows an example of a bi-crystal Burger circuit which counts 98 planes in the semiconductor and 150 in the superconductor away from the interface.~This implies that the specific interface must have at least two residual misfit dislocations in order to acquire an interfacial 3:2 domain match. We find two types of dislocations and both occur in the InAsSb close to the interface. The edge dislocations are either associated with adding (Fig. \ref{fig2} \textbf{e}) or removing (Fig. \ref{fig2} \textbf{e}) a plane. Thus, the density of dislocations is here higher than predicted by the bi-crystal Burger circuit. In the particular region of the interface shown in Fig. \ref{fig2} \textbf{d}, we find four dislocations, three additions and one removal. The color plot on the right side of Fig. \ref{fig2} \textbf{d} shows the relative bending of the semiconductor planes in relation to a reference region away from the interface. Here, going from blue, which indicates downwards bending planes, to red, which indicate upwards bending planes, is associated with a dislocation (see supplementary information S5). The color plot shows the positions of the dislocations, which can be helpful because of bi-crystal phase contrast smearing of the interface.

The fact that all dislocations are observed inside the semiconductor, rather than at the interface, indicates a lower dislocation energy in the InAsSb bulk compared to the apparently strongly bonded 3:2 domain matched interface. This differs from epitaxial InAs/Al materials where no dislocations are found in the semiconductor and the Al phase appears fully relaxed with a 5 nm thick film \cite{shabani2016two}. Since the introduction of Sb is the only difference between the two types of NWs it is reasonable to assume that the Al-Sb bonds play a major role on this strong interfacial bonding. 
We note that, even though we see dislocations appear to occur closer to the interface for Sb concentrations $x$ = 0.34 than for $x$ = 0.64, we have not been able to find a systematic trend for the dislocation depth as a function of composition.


\begin{figure}[ht!]
\vspace{0.2cm}
\includegraphics[scale=1]{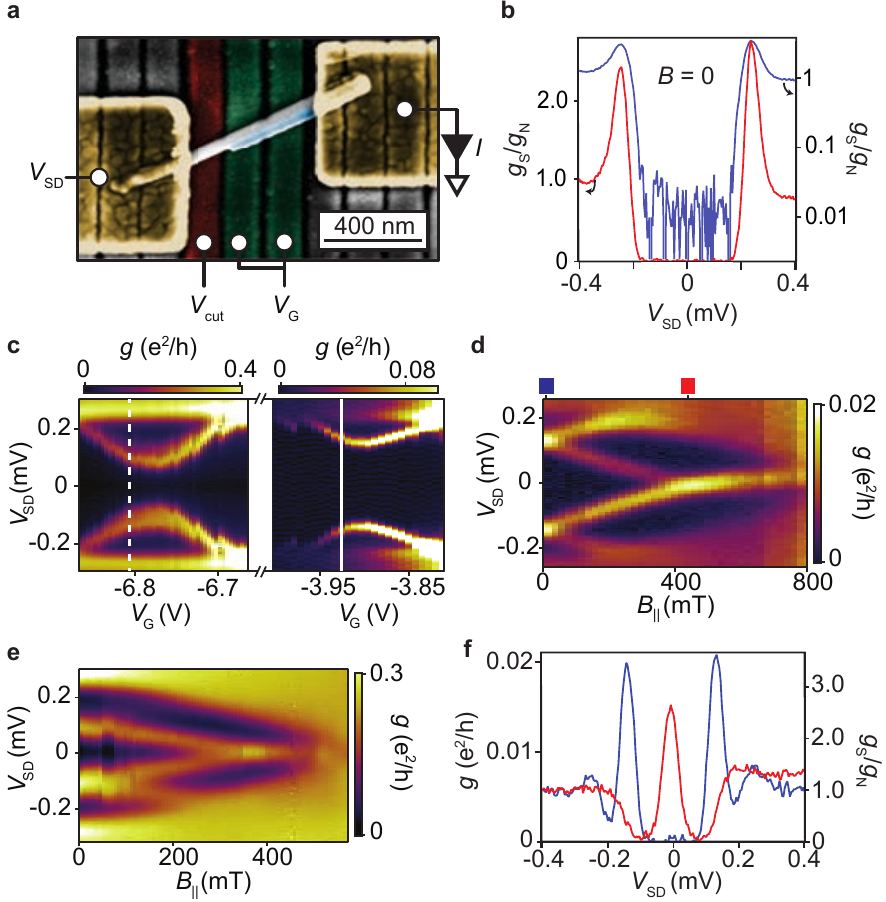}
\vspace{0.2cm}
\caption{ \textbf{NIS device.} \textbf{a}, False-colored scanning electron micrograph of a NIS device. Yellow, Ti/Au contacts; grey, InAs$_{0.8}$Sb$_{0.2}$ NW; light blue, Al shell; $V$\textsubscript{SD}, the applied voltage bias; $I$, measured current; $V$\textsubscript{G}, gate voltage controlling the chemical potential on the proximitized segment; $V$\textsubscript{cut}, gate voltage controlling the tunnel barrier height. \textbf{b}, Differential conductance, $g$, as a function of $V$\textsubscript{SD} plotted on a linear (blue trace) and logarithmic scale (red trace) showing a superconducting hard-gap with highly suppressed differential conductance within $\Delta$ = $\pm$230 $\mu$eV. \textbf{c}, $g$ as a function of $V$\textsubscript{SD} and V$_G$ showing two symmetric ABS within $\Delta$ $\sim$ 230 $\mu$eV at two different gate configurations. Unbroken and dashed lines indicate the gate configuration for \textbf{d}, and \textbf{e}, respectively. \textbf{d, e}, $g$ as a function of $V$\textsubscript{SD} and $B$\textsubscript{$\parallel$} showing the magnetic field evolution of two ABS that merge and form a zero bias peaks at $\sim$ 350 and 400 mT, respectively. \textbf{f}, $g$ as a function of $V$\textsubscript{SD} corresponding to line-cuts (blue and red square) seen in \textbf{d}, shows the zero-bias peak with strongly suppressed differential conductance symmetrically around the peak, signature of a hard topological gap.}
\label{fig3}
\end{figure}

\begin{figure}[ht!]
\vspace{0.2cm}
\includegraphics[scale=1]{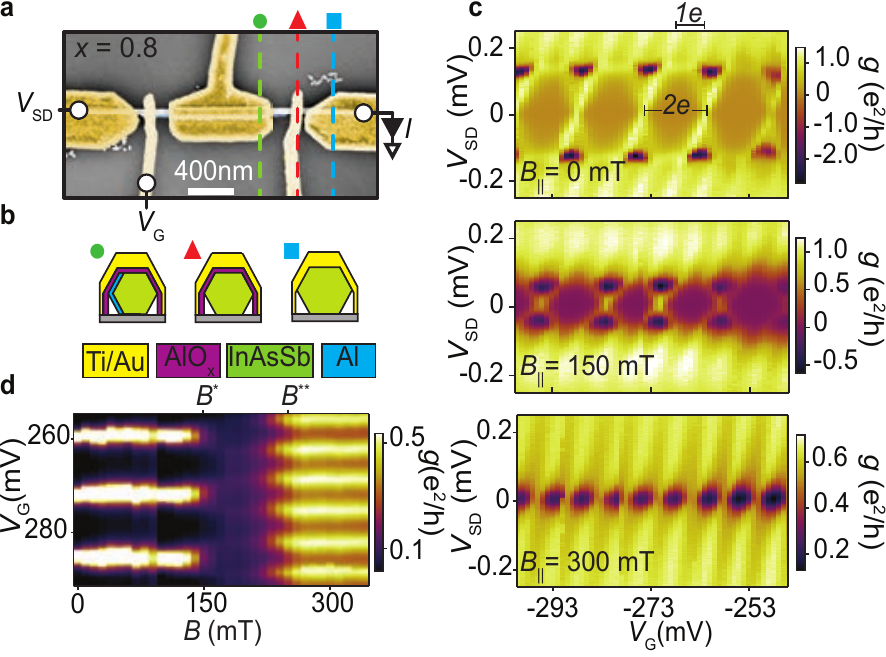}
\vspace{0.2cm}
\caption{ \textbf{NISIN device.} \textbf{a}, False-colored SEM image of the measured NISIN device. Yellow, Ti/Au contacts; green, InAs$_{0.8}$Sb$_{0.2}$ NW; blue, Al shell; $V$\textsubscript{SD}, the applied voltage bias; $I$, measured current; $V$\textsubscript{G}, gate voltage. \textbf{b}, Cartoon of the cross-sections indicated in \textbf{a} by three coloured dashed lines illustrates the top-gate material sequencing. Blue square: Ti/Au contacting the bare InAs\textsubscript{0.2}Sb\textsubscript{0.8} NW. Red triangle: Ti/Au deposited onto a few nm's of AlO\textsubscript{x} (coloured purple) on the bare NW. Green circle: Ti/Au deposited onto AlO\textsubscript{x} on the NW with 2 facet Al. \textbf{c}, differential conductance, $g$, as a function of $V$\textsubscript{G} and $V$\textsubscript{SD} showing 2$e$-spaced Coulomb diamonds within $\Delta$ $\sim$ 0.15 meV. For $V$\textsubscript{SD} $>$ 0.15 mV 1$e$ periodic Coulomb resonances are evident, verifying Cooper pair tunneling within $V$\textsubscript{SD}$\sim$ $\pm$ 0.15 mV. \textbf{d}, differential conductance at zero bias as a function of $V$\textsubscript{G} and $B_{\parallel}$. Splitting from 2$e$ to 1$e$ initiates at $B^* = B_{\parallel} = $150 mT and is evenly 1$e$ spaced at $B^{**} = B_{\parallel} = $ 250 mT, suggesting a smooth transition from 2$e$ to 1$e$.}
\label{fig4}
\end{figure}

In the following section we investigate the superconducting transport properties of the hybrid InAs\textsubscript{0.2}Sb\textsubscript{0.8}/Al material using two device geometries: a normal metal-insulator-superconductor (NIS) device and a normal metal-insulator-superconductor-insulator-normal metal (NISIN) device (Fig.~\ref{fig3}~\textbf{a} and \ref{fig4}~\textbf{a} respectively). The so-called insulator part is a segment of InAsSb where the Al is selectively etched away to enable carrier depletion by electrostatic gating. We note that we were not able to deplete carriers in devices with $x$ between 0.3-0.7, using standard gating geometries, and hence focus on devices with $x$ = 0.8.\\

Data from tunneling conductance spectra of the bottom-gated NIS device is shown in Fig. \ref{fig3} \textbf{b}. By adjusting the voltage applied on the junction, $V_{cut}$, we restrict the device in the tunneling regime with a conductance well below the conductance quantum ($g \ll 2$e$^2/h$). From the tunneling spectrum, as a function of source-drain voltage, we find the superconducting gap, $\Delta$, to be approximately $\sim$ 230 $\mu$eV. The differential conductance in the gap is suppressed by a factor of approximately $10^2$ comparing to the conductance outside of the gap, which is similar to the gap hardness reported on epitaxial InAs/Al NWs \cite{chang2015hard}, and seems to be attributed to the clean superconductor-semiconductor interface rather than specific epitaxial lattice matching. 
We note that while the critical field is comparable to what we typically find for similar Al thicknesses on InAs/Al NWs, the induced superconducting gap of $\Delta$ $\sim$ 230 $\mu$eV is significantly larger. 

In Fig. \ref{fig3} \textbf{c} differential conductance as a function of gate voltage, $V$\textsubscript{G}, and $V$\textsubscript{SD} shows two pairs of Andreev bound states (ABS) at two very different gate configurations extending out of the continuum symmetrically around $V$\textsubscript{SD} = 0. For most gate configurations at zero field the superconducting gap appears hard and empty of states. By tuning $V$\textsubscript{cut} the coupling, $\Gamma$, between the lead and the proximitized segment can be controlled, as also evident by the difference in the magnitude of differential conductance between the left and right plot. 

For gate voltages corresponding to the dashed and unbroken line-cuts in Fig. 3 \textbf{c} left and right panel, Fig. 3 \textbf{e} and 3 \textbf{d} show the respective DOS evolution in a parallel magnetic field along the NW axis. The parallel magnetic field orientation is found as similarly described in the methods section of Ref. \citen{albrecht2016exponential}. The ABS merge and pin to $V$\textsubscript{SD} = 0 at magnetic fields around $B$ $\sim$ 350 and 400 mT, respectively. This occurs at relatively low fields \cite{mourik2012signatures,deng2012anomalous,deng2016majorana} due to large effective $g$-factors of the hybrid system, in this case extracted to be on the order of 10. 
Like the overall conductance, also the bound state conductance and ZBP height depends strongly on $V_{cut}$. For the ZPBs it was not possible exceed 0.3 $\cdot$ 2e$^2/h$ by opening $V_{cut}$ before the gap softened and the overall conductance increased. This indicates that the cutter region where the Al is etched away does not fulfill resonant tunneling. Ideal tunneling was recently observed in two-dimensional hybrid materials with a thin etch stop layer between the Al and the InAs, and also in hybrid nanowires with 'shadowed' junctions \cite{Fabrizio2e2h, Leo2e2h}. The ZBPs prevail until the closing of the superconducting gap at critical fields of the Al at $B$\textsubscript{c} $\sim$ 600-800 mT. The observed critical fields are low compared to similar experiments conducted on InAs NWs proximitized by Al which yielded critical fields of $B$\textsubscript{c} $\sim$ 2 T with a film thickness of 7 nm \cite{deng2016majorana}. We attribute the lower critical fields to the average Al film thickness which in our case is about 15 nm \cite{meservey1971properties}. The blue and red square in Fig. 3 \textbf{d} correspond to the two line-traces in Fig. 3 \textbf{f}, where the normalized differential conductance as a function of $V$\textsubscript{SD} at $B_{\parallel}$ = 0 and 430 mT is plotted. We find that the gap stays hard around the ZBP at $B$\textsubscript{$\parallel$} = 430 mT, with a gap of the order of $\Delta$. Extracted $g$-factors of the hybrid system at various gate configurations are generally extracted to be between 10 and 20 as shown in supplementary information S6.

Additional transport measurements were performed on the same batch of InAs\textsubscript{0.2}Sb\textsubscript{0.8}/Al NWs using a normal metal-insulator-superconductor-insulator-normal metal (NISIN) geometry. In Fig. \ref{fig4} \textbf{a} we show a scanning electron micrograph of a measured NISIN device with a superconductor island length of $L$ = 800 nm. In this device the gates are realized as top gates using AlO\textsubscript{x} as dielectric. Schematics of the cross-sections of the different material stacking sequences of the device is seen in Fig. 4 \textbf{b}.  


Differential conductance as a function of $V$\textsubscript{SD} and $V$\textsubscript{G} at different magnitudes of parallel magnetic fields are shown in Fig. \ref{fig4} \textbf{c}. Evenly spaced Coulomb diamonds are observed as well as differential conductance resonances above the superconducting gap. At $B_{\parallel}$= 0 and $V$\textsubscript{SD} = 0 each Coulomb resonance occurs at double the gate voltage separation compared to the period observed above the superconducting gap. The period doubling within the superconducting gap implies Cooper pair tunneling of charge 2$e$ as opposed to the 1$e$ periodicity above the superconducting gap. As previously reported from experiments on superconductor islands \cite{higginbotham2015parity, hekking1993coulomb, hergenrother1994charge}, we see regions with strong negative differential conductance (NDC) at bias voltages where the 2$e$ to 1$e$ transitions occur. At $B$ = 150 mT the 2$e$ periodic Coulomb diamonds are still observed, however with a decreased superconducting gap and magnitude of the NDC. At $B$ = 300 mT, well below the critical field, $B$\textsubscript{c} of the Al, the Coulomb resonances exhibit a 1$e$ periodicity. This is consistent with the energy of the odd charge state being lowered sufficiently by the Zeeman energy in order for the 1$e$ periodicity to become dominant \cite{albrecht2016exponential}. We speculate that the lack of visible oscillations in the even-odd energy difference,~as observed in Ref.~\citen{albrecht2016exponential} could be due to the combination of a relative large island size and high SOI for this material, producing a minimal overlap between the Majorana wave functions. Tunneling from the bound state to the Bardeen-Cooper-Schrieffer (BCS) continuum in the Al gives rise to NDC due to transport blockade by electron/hole excitations.~When the quasi-particle relaxes from the continuum back into the bound state and escapes to the leads the transport blockade is lifted. This way large NDC indicates a relative long quasi-particle relaxation times as explained in Ref. \citen{higginbotham2015parity}.

Figure 4 \textbf{d} shows the evolution of the zero-bias Coulomb peaks in a parallel magnetic field. A distinct 2$e$ peak spacing is evident until $B$ = 150 mT where the splitting initiates. Noticeably, the intensity of the zero-bias resonances decrease drastically as it evolves into a 1$e$ periodicity between $B$\textsuperscript{*} and $B$\textsuperscript{**}. From $B$ = 250 mT the evenly spaced Coulomb resonances regain intensity but with a 1$e$ evenly spaced periodicity and no signatures of even-odd intensity variance - again consistent with transport through topologically protected MZM \cite{fu2010electron, hutzen2012majorana}. The splitting is completed well below the closing of the induced superconducting gap, which is spectroscopically extracted to be between $B$\textsubscript{c} $\sim$ 600-800 mT (depending on gate configuration). For example, residual superconductivity is still clearly visible at $B$ = 300 mT, as seen in Fig. \ref{fig4} \textbf{c}. 

\begin{figure}[ht!]
\vspace{0.2cm}
\includegraphics[scale=1]{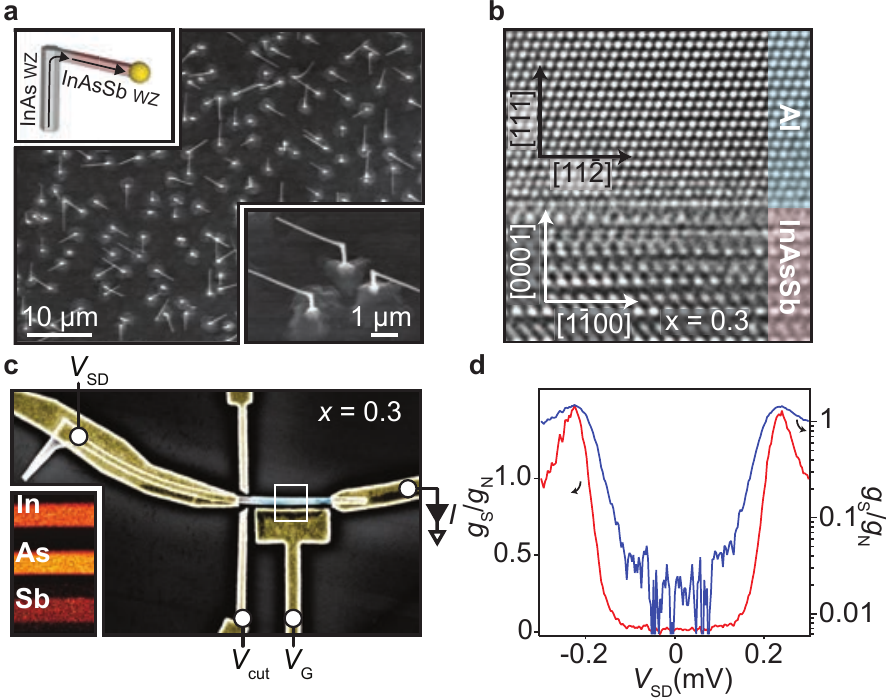}
\vspace{0.2cm}
\caption{\textbf{InAs$_{0.7}$Sb$_{0.3}$ WZ NW grown along [1$\bar{1}$00].} \textbf{a}, Top-view SEM micrograph of the NWs grown in the six direction parallel to the substrate. Inset show a $45^{\circ}$ tilted SEM micrograph of three kinked structures. \textbf{b}, HR-TEM micrograph of the InAsSb/Al interface shows a coherent epitaxial crystal match. The black and white arrows indicate the orientations of the hybrid system, where a 3:2 domain match with no rotation is observed. \textbf{c} False-colored SEM micrograph of a NIS device composed of normal metal leads connected to a InAs$_{0.7}$Sb$_{0.3}$ WZ NW grown along the [1$\bar{1}$00]-direction. Yellow, Ti/Au contacts; non-colored, InAs$_{0.7}$Sb$_{0.3}$ WZ NW; light blue, Al shell; $V$\textsubscript{SD}, these applied voltage bias; $I$, measured current; $V$\textsubscript{G}, gate voltage controlling the chemical potential on the proximitized segment; $V$\textsubscript{cut}, gate voltage controlling the tunnel barrier height. Inset shows normalized EDX intensities for In, As and Sb from a region of the NW shown by the white box.\textbf{d}, Differential conductance, $g$, as a function of $V$\textsubscript{SD} plotted on a linear (red trace) and logarithmic scale (blue trace) showing a superconducting hard-gap with highly suppressed differential conductance within $\Delta$ = $\pm$225 $\mu$eV.}
\label{fig5}
\end{figure}

As discussed above, we find that ZB InAs$_{1-x}$Sb$_{x}$ NWs exhibit a weak electrostatic gate response for a range of compositions. The electron affinity of the InAsSb material crucially influences the filling of the conduction band, and therefore how easily the NW can be gated and possibly depleted. As the electron affinity of III-V materials typically is higher in ZB than WZ \cite{yeh1992zinc}, we expect that InAsSb NWs with a WZ crystal structure could improve gatability. Since it is challenging to form WZ phases in NWs containing Sb, because of a large bulk cohesive energy difference between ZB and WZ \cite{dubrovskii2014nucleation,ghalamestani2016can}, we grow from a WZ basis by initiating growth from the (1$\bar{1}$00) sidefacets of [0001] InAs WZ NWs, as explained in Ref. \citen{Krogstrup2009, krizek2017growth}. This is illustrated in the inset in Fig. \ref{fig5} \textbf{a}. As long as the NWs grow layer-by-layer in the [1$\bar{1}$00]-direction, the crystal structure is locked in the parent WZ crystal basis, and can in this case not change to ZB without introducing a high energy incoherent interface. After initiating the [1$\bar{1}$00] InAs growth, we gradually introduce Sb over a segment of $\sim$ 1 \textmu m to avoid secondary kinking, and continue the InAsSb growth for $\sim$ 5 \textmu m. We terminate the growth with an epitaxially grown shell of Al on the top facet. The NWs as-grown are seen in Fig. \ref{fig5} \textbf{a}. A high-resolution TEM micrograph of the bi-crystal interfacial match between the WZ InAs\textsubscript{0.7}Sb\textsubscript{0.3} NW and the Al shell is seen in Fig. \ref{fig5} \textbf{b}. Similarly to the ZB hybrids presented in Fig. \ref{fig2}, we find a well-defined and epitaxial interface with a 3:2 domain match along the growth direction. In contrast to the similar InAs/Al NWs \cite{krogstrup2015epitaxy}, the residual mismatch along the [1$\bar{1}$00]-direction is negative and thus bending the NW upwards as evident in Fig. \ref{fig5} \textbf{c}. In Fig. \ref{fig5} \textbf{c} we show a false-colored SEM micrograph of the InAs\textsubscript{0.7}Sb\textsubscript{0.3} WZ NW connected by normal-metal leads. The device geometry is similar to the geometry shown in Fig. \ref{fig3}, except side-gates are used instead of bottom gates. The inset shows an EDX map with normalized intensities for In, As and Sb, performed on the last part of a kinked NW as illustrated by the white square. This verifies a constant composition along the growth direction with no detectable Sb shell formation and a quantified composition of $x$ = 0.3. In Fig. \ref{fig5} \textbf{d} the left y-axis shows normalized differential conductance as a function of $V$\textsubscript{SD} displaying a hard-gap profile in the DOS. The right y-axis shows the same normalized differential conductance plotted on a logarithmic scale where the differential conductance within $\Delta$ = 0.225 meV is suppressed by a factor of $\sim$ 80. These findings are comparable to transport measurements performed on the ZB InAs\textsubscript{0.2}Sb\textsubscript{0.8} NWs. Assuming that the WZ electron affinity as a function of composition follows that of ZB, a composition of $x$ = 0.3 would have an effective maximum \cite{webster2015measurement}, thus it is likely that all compositions of WZ InAs$_{1-x}$Sb$_{x}$ NWs would be depletable simply by applying standard side-gates. Additional advantages to WZ NWs grown in the [1$\bar{1}$00] direction could be a contribution to the SOI from the Dresselhaus spin-orbit term that is linear in momentum, arising solely due to asymmetry in the crystal structure \cite{sand2017}. 

In conclusion, we show from WAL measurements that the SOI as a function of composition in InAs$_{1-x}$Sb$_{x}$ NWs exhibits a non-monotonic behavior with a maximum in the Rashba spin-orbit coefficient at $x$ $\sim$ 0.5. By growing Al shells in-situ on InAs$_{1-x}$Sb$_{x}$ NWs with varied compositions we find an epitaxial relationship with a specific low energy interfacial domain of 3:2 in the axial direction for all compositions. Axial and radial Al rotations combined with interfacial edge dislocations seem to reduce the contribution to the excess energy. Coulomb charging and tunneling measurements on InAs$_{0.2}$Sb$_{0.8}$/Al hybrid NWs show hard gap induced superconductivity, as well as data consistent with topological superconductivity. 

From the tunneling experiments we find a ZBP protected by a hard superconducting gap.~Additionally, from the Coulomb charging experiments we find strong NDC indicating long quasi-particle poisoning times. 

In the end we present a method to grow InAsSb NWs with WZ structure, which shows an improved electrostatic response for InAs$_{0.7}$Sb$_{0.3}$ NWs grown along the [1$\bar{1}$00]-direction.~The strong Rashba SOI, the good electrostatic response, the well defined epitaxial superconductor-semiconductor match and the potential addition to an even higher SOI due to a potential Dresselhaus contribution, are all ingredients that make WZ InAs$_{0.5}$Sb$_{0.5}$/Al hybrid NWs a promising material for further studies on topological superconductivity.

\section{Acknowledgement}
This project was funded Microsoft Station Q, the Danish National Science Research Foundation, the Carlsberg Foundation, the Villum Foundation and European Research Council (ERC) supported the research under grant agreement No.716655 (\textit{HEMs-DAM}). We thank Esben Bork Hansen for fruitful discussions, and Claus B. S\o rensen and Shivendra Upadhyay for technical assistance. 

\section{Author Contributions}

JES, TSJ, ANG, DS, JY and MvS fabricated the devices and carried out transport measurements with support and input from MD, TSJ, JN, CMM, PK. PK and TK developed the material growth and analysis hereof. EJ and TK performed the TEM characterization. All authors contributed to interpreting the data. The manuscript was written by JES, TK, ANG and PK with input from all other authors.

\section{Competing financial interests}
The authors declare no competing financial interests.

\section{Methods}

All nanowires are grown by MBE on InAs(111)B substrates and catalyzed by Au via the vapor-liquid-solid method at a substrate temperature of $T$\textsubscript{sub} = 420 $^{\circ}$C. Following the process described by Ref.~\cite{caroff_InSb} the NW growth is initiated with a stem of InAs along the [0001] direction, using an In flux corresponding to a planar InAs growth rate of 0.5 $\mu$m/hr and a calibrated As$_4$/In flux ratio of 14. Subsequently, Sb is introduced while keeping the total group V flux constant. For Sb$_x$/As$_4$ flux ratios of up to $\sim$ 0.1, the crystal structure changes from WZ to a periodic twinning zincblende phase, as reported by Ref.~\citen{xu2012faceting} and shown in supplementary information S1. By increasing the Sb concentration the crystal structure becomes pure ZB without twin planes. The constant total group V flux keeps the NW diameter almost constant as a function of composition, all the way to pure InSb. For the axial heterostructure NWs presented in Fig.~\ref{fig1} each segment is grown for 180 seconds with a constant group V/III flux ratio of 14.~See Supplementary Information S1 and S2 for a more detailed discussion on the NW growth and structure.
The Al shell was grown at -30 $^{\circ}$C on 2 facets of the hexagonal cross-section with a flux corresponding to a AlAs planar growth rate of 1 monolayers per second. \\

For the device fabrication, the NWs were deposited onto doped Si substrates with a SiO$_2$ thickness of 100 nm, either by dry deposition, use of the tip of a cleanroom wipe or by selecting individual NWs using of a micro-manipulator needle with a diameter of $\sim$ 250 nm under an optical microscope. Ohmic contacts to the NWs was ensured using RF ion milling at 25 W for 3 m 20 s before evaporating 5 nm Ti and 120 nm Au. In case of NWs with an Al shell growth, the shell was selectively removed by defining etching windows via electron beam lithography and etching in Transene D for 10 s at 50 C. See supplementary information S3 for detailed recipes. \\  

\bibliography{ref2}
\end{document}


\section*{S1 - Growth of InAs\textsubscript{1-x}Sb\textsubscript{x} nanowires}

\begin{figure}[htb!]
\centering
	\includegraphics[scale = 1]{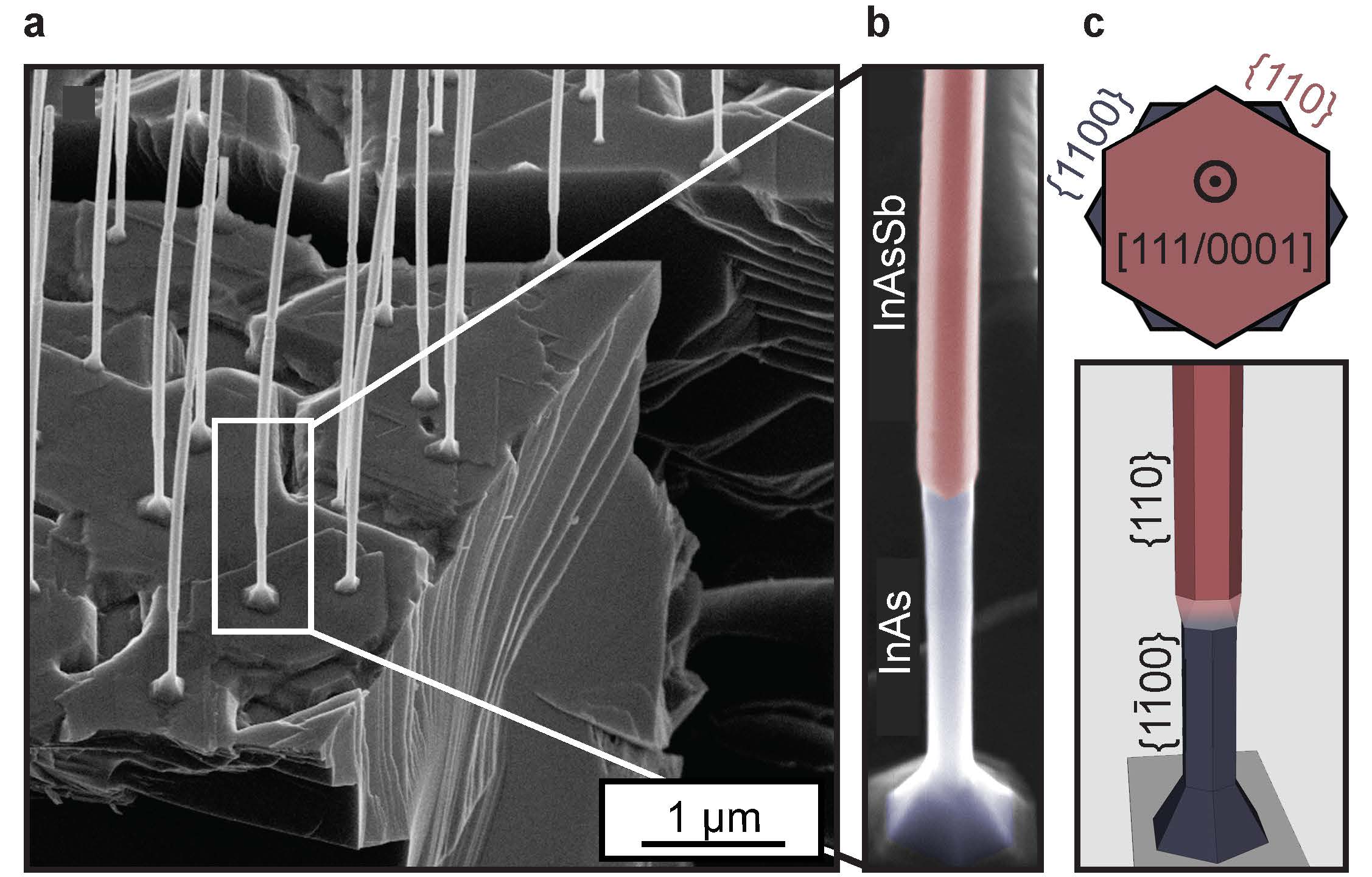}
	\caption{\textbf{InAs/InAsSb facetting.} \textbf{a}, A scanning electron microscope (SEM) overview micrograph of the cleaved growth wafer. \textbf{b}, False colored SEM micrograph of individual NW. Blue indicates InAs and red indicates InAsSb. Pronounced faceting can be seen at the root of the InAs segment, and on the InAsSb segment. A $30^{\circ}$ rotation between the two facets is seen. \textbf{c}, Illustration of an axial view (on top) and radial view (below). The facets are found to be $\{ 1 \bar{1}00 \}$ for pure InAs and $\{1\bar{1}0 \}$ for InAsSb.}
	\label{facetrot}
\end{figure}

Under given nanowire (NW) growth conditions (see method section in main text), the InAs wurtzite (WZ) stem forms $\{ 1 \bar{1}00 \}$ facets and the InAsSb zincblende (ZB) structure form $\{1\bar{1}0 \}$ facets. The two family of facets are rotated 30$^{\circ}$ with respect to each other, as shown and illustrated in Fig. \ref{facetrot} \textbf{a-c}. The rotation takes place as the structure switches to ZB when Sb is introduced.\\

\begin{figure}[htb!]
\centering
	\includegraphics[scale=1]{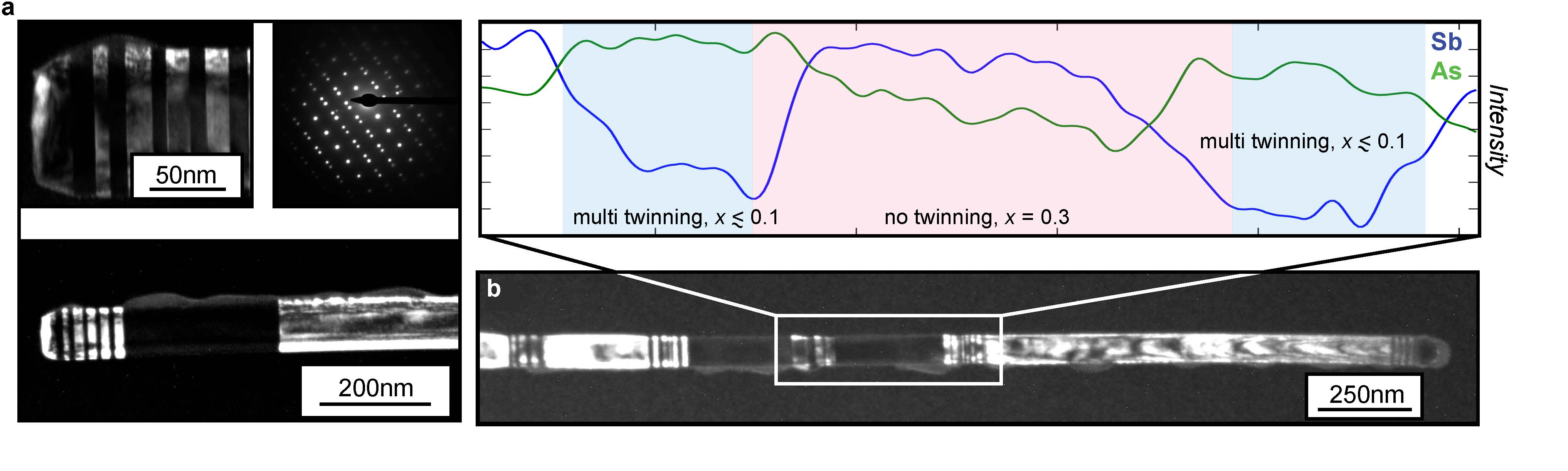}
	\caption{\textbf{Multi-twinning sequences.} \textbf{a}, Dark-field TEM micrographs and selective area electron diffraction (SAED) pattern of the multi-twinning sequence at the transition from pure InAs to InAsSb (InAs not visible here). The zone axis of the SAED pattern is [$1\bar{1}0$] of the two twin orientations. The dark-field TEM micrograph is acquired by selecting one of the ZB twinning segments. \textbf{b}, Dark-field TEM micrograph of an InAsSb axial heterostructure NW with multi-twinning segments at low Sb concentrations, as shown in the zoom-in in the (non-quantified) EDX line scan.}
	\label{size}
\end{figure}

The structural transition from pure InAs (with WZ structure) to InAsSb (with ZB structure) at low Sb concentrations $\lesssim$ 0.1, before reaching pure ZB structure at higher concentrations, consists of a multi-twining phase as observed by Ref. [\citen{potts2015twinning}]. This is shown in Fig. \ref{size} \textbf{a}, where NWs are grown with an alternating composition of $x$ $\lesssim$ 0.1 and $x$ = 0.3. The dark and bright contrast corresponds to the two twin orientations of the ZB structure. In Fig. \ref{size} \textbf{b} the same twinning sequences are observed in the dark-field transmission electron microscope (TEM) micrograph of an InAs\textsubscript{1-x}Sb\textsubscript{x} axially segmented heterostructure. The multi-twining phase appears to be related to composition rather than to the actual transition between different compositions.\\


\begin{figure}[htb!]
\centering
	\includegraphics[scale = 1]{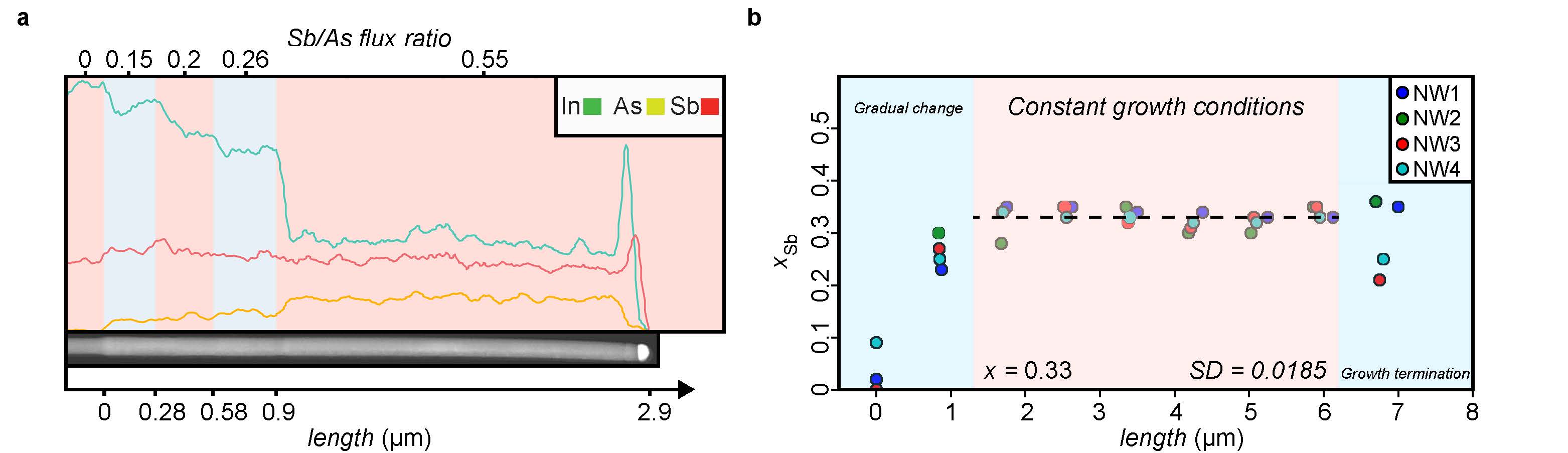}
	\caption{ \textbf{EDX line scans of InAsSb NWs.} \textbf{a}, EDX intensity line scan of In, As and Sb on the HAADF-STEM image of the NW below the plot. Blue and pink background represents the different concentration steps. The axis above the graph represents the group V beam flux ratio, and the axis below shows the length along the NW. \textbf{b}, Manual line scans of the quantified data as a function of position along the NW for 4 different NWs from the same growth. Assuming constant composition the standard deviation is $SD$ = 0.0185, which we expect to be less than uncertainties related to EDX indicating stable NW compositions at $x$ = 0.33 for $\sim$ 5 micrometers under constant growth parameters.} \label{inassb_1}
\end{figure}

As the condition for surface diffusion changes as NWs grows longer, it is not obvious that the composition will stay constant under constant fluxes. However, as we will show below, it is possible to obtain a constant composition (within the energy-dispersive x-ray spectroscopy (EDX) detection uncertainty) over several microns. In Fig. \ref{inassb_1} \textbf{a}, we show a high-angle annular dark-field (HAADF) scanning TEM micrograph of an InAs$_{1-x}$Sb$_x$ NW along with an EDX intensity line scan, where the compositional change is due to the abrupt changes of group V flux ratios. In Fig. \ref{inassb_1} \textbf{b} the Sb molar fraction from 4 different NWs with different diameters are plotted as a function of the position along the length of the NW. The linescan is initiated from the pure InAs until it saturates at the requested concentration. The quantified EDX data in Fig. \ref{inassb_1} \textbf{b} shows the constant Sb molar fraction over $\sim$ 5 $\mu$m, until the growth is terminated and the cool down process is initiated.


\newpage

\section*{S2 - Axially segmented InAsSb heterostructures.}
\begin{figure}[htb!]
\centering
	\includegraphics[scale = 1]{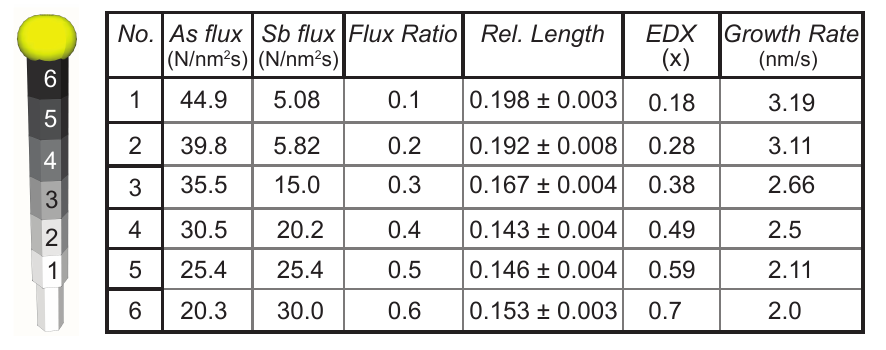}
	\caption{\textbf{Schematic and table of growth details on the axial heterostructure NW.}}
	\label{grad_1}
\end{figure}

The axial heterostructure NW presented in Fig. 1 is grown with the parameters presented in Fig. \ref{grad_1}. The fluxes and ratios for each individual segment is presented alongside the composition measured by EDX. For increasingly higher Sb ratios the growth rate seems to decrease. The relative lengths of each segment are used for exact placement of the Au leads on the compositional transitions in the nanowire as described in S3. 

\newpage

\section*{S3 - Device Fabrication\label{devicefab}.}

This section regards the fabrication process of the presented side gated, bottom gated and top gated devices presented in Fig. 1 \textbf{b}, 3 \textbf{a}, 4 \textbf{a} and 5 \textbf{c}. 

\subparagraph{Side gated devices}

\begin{itemize}
\item 1) Pick up and deposit NWs onto highly doped Si chip with 100 nm thermal oxide by micro needle manipulator.
\item 2) Spin on PMMA A4 950 at 4000 RPM for 45 s and bake at 185 $^{\circ}$C for 2 min. Repeat to make double layer.
\item 3) The placement of gates and contacts is designed according to the relative lengths extracted for each axial segment of differing composition. The initial InAs stem and cooling neck of the deposited NW is used to verify the true length of the NW at hand by SEM and the relative length (found in S2) is used to determine the actual length of each individual segment. E-beam lithography is used to write the contacts and gates.  
\item 4) Develop in MIBK/IPA (1:3) for 1 min, rinse in IPA for 30 s and oxygen plasma ash for 1 min at 0.3 mbar and 3.2 sccm. 
\item 5) RF argon mill for 3 min at 25 W with an Ar flow of 30 sccm, before e-beam evaporating 5nm Ti and 120 nm Au. 
\item 6) Lift-off in warm acetone (55 $^{\circ}$C) for 60 min. 
\item 7) Quick rinse in acetone followed by 1 min in IPA and 1 min in ultra pure water.

\end{itemize}
\subparagraph{Bottom gated devices.}

\begin{itemize}
\item 1) Spin PMMA A2 950 at 4000 RPM for 45 s on highly doped Si chip with 100 nm thermal oxide. Bake at 185 $^{\circ}$C for 2 min.
\item 2) Write gate electrodes with e-beam lithography system.
\item 3) Develop in MIBK/IPA (1:3) for 1 min, rinse in IPA for 30 s and oxygen plasma ash for 1 min at 0.3 mbar and 3.2 sccm. 
\item 4) Deposit e-beam evaporated Ti/Au (5nm/20nm). 
\item 5) Lift off in warm acetone (55 $^{\circ}$C) for 60 min.
\item 6) Spin on co-polymer 9 \% at 4000 RPM for 45 s and bake at 185 $^{\circ}$C for 2 min. Repeat step with PMMA A4 950 to create double layer. 
\item 7) Write windows positioned over the bottom electrodes by e-beam lithography. 
\item 8) Develop in MIBK/IPA (1:3) for 1 min, rinse in IPA for 30 s and oxygen plasma ash for 1 min at 0.3 mbar and 3.2 sccm.
\item 10) Grow 12 nm of HfO\textsubscript{x} by atomic layer deposition (ALD).
\item 11) Lift-off in warm acetone (40 $^{\circ}$C) while sonicating at high power and low frequency for 10 min in order to knock down HfO\textsubscript{x} build-up. 
\item 12) Pick up and deposit NWs onto ALD covered bottom electrodes by micro needle manipulator.
\item 13) Spin single layer of PMMA A4 950 at 4000 RPM for 45 s Bake at 185 $^{\circ}$C for 2 min. 
\item 14) Write etching windows with e-beam lithography system.
\item 15) Develop in MIBK/IPA (1:3) for 1 min, rinse in IPA for 30 s and oxygen plasma ash for 1 min at 0.3 mbar and 3.2 sccm. 
\item 16) In a beaker with closed lid heat Al etch transene D in a heat-bath to 50 $^{\circ}$C and let sit for 15 min, to stabilize temperature. Etch in the heated solution while swirling for 10 s. Clean in 3 beakers of ultra pure water.
\item 17) Strip resist in warm acetone (55 $^{\circ}$C) for 10 min and oxygen plasma ash for 1 min at 0.3 mbar and 3.2 sccm.  
\item 17) Spin on PMMA A4 950 at 4000 RPM for 45 s bake at 185 $^{\circ}$C for 2 min. Repeat to create double layer. 
\item 18) Write contacts and connecting lines to the bottom electrodes with e-beam lithography system.
\item 19) Develop in MIBK/IPA (1:3) for 1 min, rinse in IPA for 30 s and oxygen plasma ash for 1 min at 0.3 mbar and 3.2 sccm. 
\item 20) RF argon mill for 3 min and 20 s at 25 W before evaporating 5nm Ti and 120 nm Au. 
\item 21) Lift-off in warm acetone (55 $^{\circ}$C) for 60 min.
\item 22) Quick rinse in acetone and 1 min in IPA and 1 min ultra pure water. 
\end{itemize}

\subparagraph{Top gated devices.}

\begin{itemize}
\item 1) Pick up and deposit NWs onto highly doped Si chip with 100 nm thermal oxide by micro needle manipulator.
\item 2) Spin on PMMA A4 950 at 4000 RPM for 45 s bake at 185 $^{\circ}$C for 2 min. Repeat to create double layer.
\item 3) Write contacts with e-beam lithography system.
\item 4) Develop in MIBK/IPA (1:3) for 1 min, rinse in IPA for 30 s and oxygen plasma ash for 1 min at 0.3 mbar and 3.2 sccm. 
\item 5) In a beaker with closed lid heat Al etch transene D in a heat-bath to 50 $^{\circ}$C and let sit for 15 min, to stabilize temperature. Etch in the heated solution while swirling for 10 s. Clean in 3 beakers of ultra pure water.
\item 6) RF argon mill for 3 min  20 s at 25 W before evaporating 5nm Ti and 120 nm Au. 
\item 7) Lift-off in warm acetone (55 $^{\circ}$C) for 60 min. 
\item 8) Quick rinse in fresh acetone and 1 min in IPA and 1 min ultra pure water. 
\item 9) Spin on PMMA A4 950 at 4000 RPM for 45 s bake at 185 $^{\circ}$C for 2 min. Repeat to create double layer. 
\item 10) Write top gates with e-beam lithography system.
\item 11) Develop in MIBK/IPA (1:3), rinse in IPA for 30 s oxygen plasma ash for 1 min  with 
\item 12) Deposit 2 nm of Al expose to atmospheric air for 5 min Repeat 5 times before depositing 5 nm of Ti and 100 nm Au. 
\item 13) Lift-off in warm acetone (55 $^{\circ}$C) for 60 min 
\item 14) Quick rinse in acetone and 1 min in IPA and 1 min ultra pure water. 
\end{itemize}

\newpage

\section*{S4 - Phase coherence lengths.}

\begin{figure}[htb!]
\centering
	\includegraphics[scale=1]{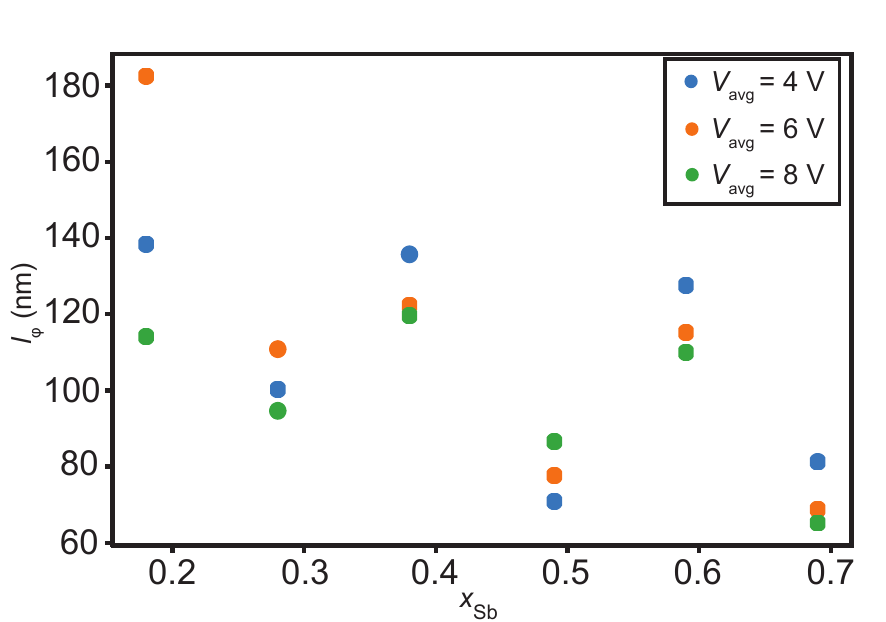}
	\caption{\textbf{Phase coherence length parameters.} \textbf{a}, Phase coherence length as a function of composition plotted for three averaging voltages, $V$\textsubscript{avg} = 4, 6 and 8 V, and $C$ = 3. }
	\label{lphi}
\end{figure}

Besides the fitting parameter $l$\textsubscript{SO} shown in the main text and the fixed NW diameter $D$ = 100 nm, the other fitting parameter used in eq. 1 in the main text is $l_{\phi}$ which is a measure of the phase coherence length. The phase coherence parameter is plotted as a function of composition as seen in Fig. \ref{lphi} \textbf{a}. These values are extracted at different averaging voltages, $V$\textsubscript{avg}.

The phase coherence lengths extracted in this work range from 60-180 nm and are a little smaller or on the same order as found by Ref. [\citen{hansen2005spin, roulleau2010suppression, hernandez2010spin}].

\section*{S5 - InAsSb/Al bi-crystal relations.}

\begin{figure}[htb!]
\centering
	\includegraphics[scale=1]{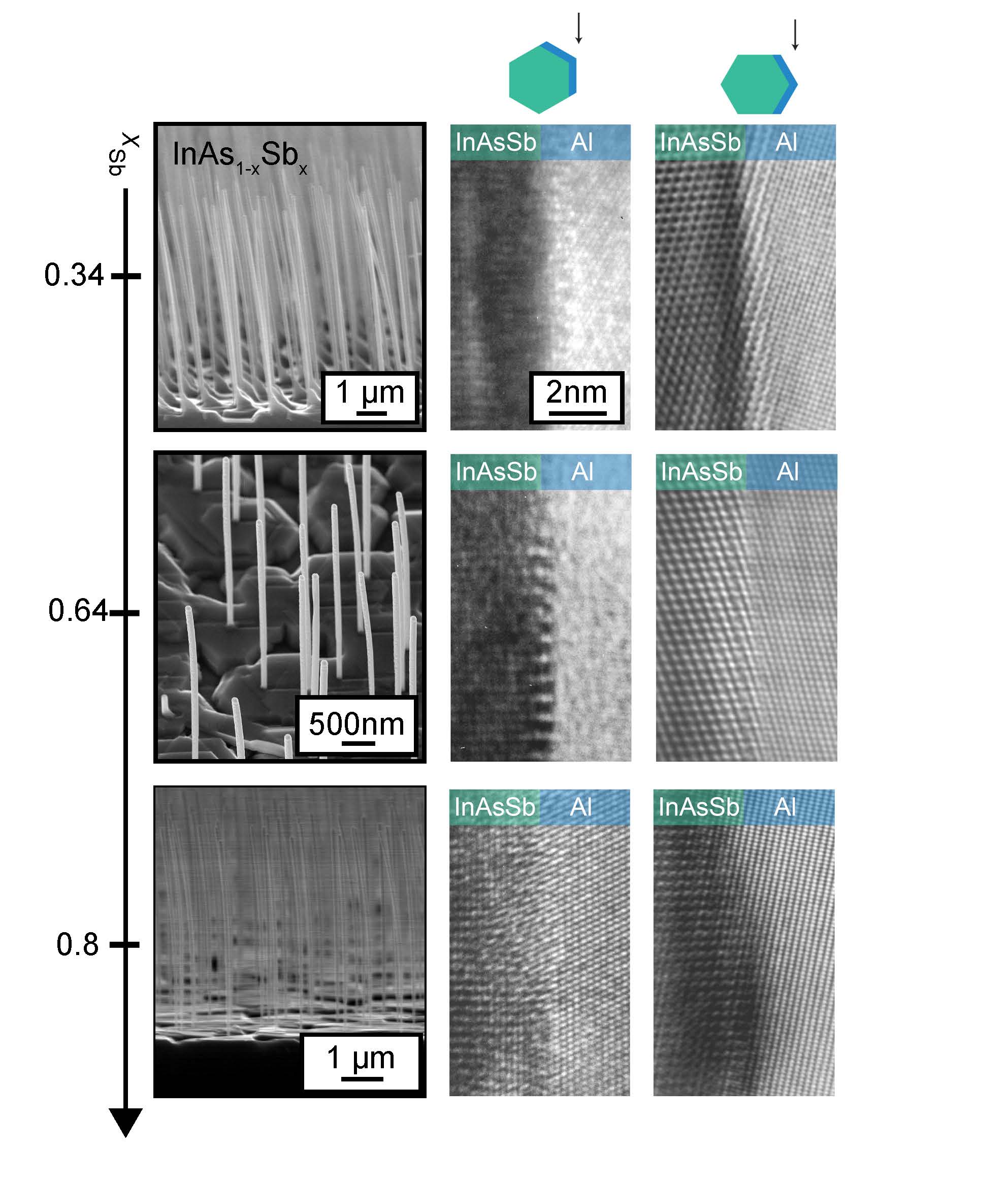}
	\caption{\textbf{SEM and TEM micrographs of InAs\textsubscript{1-x}Sb\textsubscript{x} hybrid NWs.} Descending rows correspond to increasing Sb concentration beginning at $x$ = 0.34 and ending at 0.8. First column shows an overview SEM image of the growth substrate. Second and third columns show HR-TEM micrographs of the bi-crystal matching along the semiconductor $\{11\bar{2}\}$ and $\{1 \bar{1}0\}$ crystal orientations.}
	\label{inassb_Al_1}
\end{figure}

Three different growths with a composition of $x$ = 0.34, 0.64 and 0.80 are shown in Fig. \ref{inassb_Al_1}. Column one shows an SEM overview micrograph of each growth. The other two columns show HR-TEM micrographs of the semiconductor/superconductor interfaces, along an indicated facet and corner. We find that an epitaxial Al matching on InAs\textsubscript{1-x}Sb\textsubscript{x} is feasible on all investigated InAs\textsubscript{1-x}Sb\textsubscript{x} compositions.

The transverse match of the bi-crystal interface is estimated by assuming the Al grain growth to be driven by the minimization of free energy under the constrain of kinetic barriers. In this case, the growth rate can, under the assumption of negligible desorption, from Ref. [\citen{krogstrup2015epitaxy}] be expressed as 

$$
\Delta \Gamma \propto \exp\Big( -\frac{\delta g}{k_B T}\Big) \Big(\exp \Big( \frac{\delta \mu}{k_B T} \Big) -1\Big)
$$

where $\delta g$ is the kinetic barrier for transition and $\delta \mu$ is the chemical potential. As Al is solely found in face centered cubic crystal structure, we need to look at the terms that contribute to the thermodynamic excess of the chemical potential. 

$$\label{energies}
\delta \mu = \delta_{VS} + \delta \mu_{GB} + \delta \mu_{SS}  + \delta \mu_S
$$

Here $\delta_{VS}$ is the vapor/solid Al out-of-plane surface energy and $\delta \mu_{GB}$ is the grain boundary energy. The surface energy can in principle be calculated by density functional theory\cite{fiolhais2003extraction, vitos1998surface} and we will in the later sections, use that 

$$
\delta_{VS}^{111} < \delta_{VS}^{100} < \delta_{VS}^{110}.
$$

$\delta \mu_{SS}$ is the semiconductor/superconductor interface energy and $\delta \mu_{S}$ is the interfacial strain energy. Unfortunately, we do not have any theoretical calculations of the interface energy to date and therefore assume that the transverse match should be a low index as the axial match. In broad terms we describe the contribution of interfacial strain to the grain strain energy as the interfacial strain energy density\cite{krogstrup2015epitaxy} as 

$$
\delta \mu_S = \frac{\boldmath(S) \epsilon}{(1-\nu)}
$$
where
\begin{equation} \label{together}
\epsilon^2 = \epsilon_{ax}^2 + \epsilon_{tr}^2 +2 \nu \epsilon_{ax}\epsilon_{tr}.
\end{equation}
Here $\epsilon_{ax}$ and $\epsilon_{tr}$ is the strain in the axial and transverse direction. We denote $\nu$ as the Poisson ratio for Al. As we know the specific bi-crystal match for the different compositions and assume that the contribution to the overall excess energy from the surface energy and contribution from grain boundaries is interdependent of the transverse match we can by minimizing eq. \ref{together} for all low index match estimate the transverse match.  

\begin{figure}[htb!]
\centering
	\includegraphics[scale = 1]{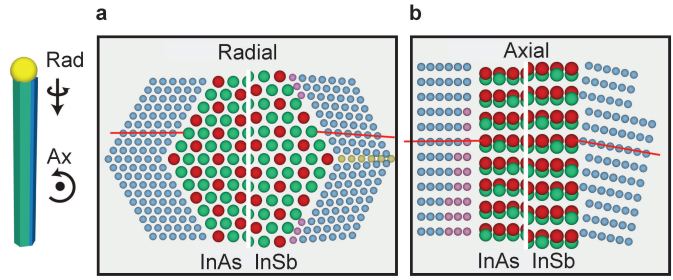}
	\caption{\textbf{Al rotation on the axial and radial axis.} \textbf{a}, Comparison between the radial rotation of the Al phase on InSb and the Al phase rotation on InAs. \textbf{b}, Similar as \textbf{a} but along the axial rotation.}
	\label{rotation}
\end{figure}

As described in the main text, the Al grains trends to rotate as the interfacial strain increases. This is due to minimization the interfacial strain energy. As seen in Fig. \ref{rotation} \textbf{a} the Al phase acquires radial and axial rotations that rotate the Al phase around the semiconductor $[111]$ and $[11\bar{2}]$, respectively. These rotations minimize the strain energy due to axial and transverse strain. An axial rotation will effectively reduce the interfacial strain in the axial direction and an radial rotation will reduce the transverse strain. 

As the rotations always are observed in discrete rotations we speculate if the discreteness of the rotations is due to specific low interface energies. These rotations could be caused by point defects at the interface, as marked in \ref{rotation} with purple. The point defects can either be an addition to the interface atoms, as marked on the radial case, or a removal, as marked on the axial case.






\newpage

\section*{S7 - Microscopic calculation of the magnetic dephasing length.}

\begin{figure}[htb!]
\centering
	\includegraphics[scale = 0.5]{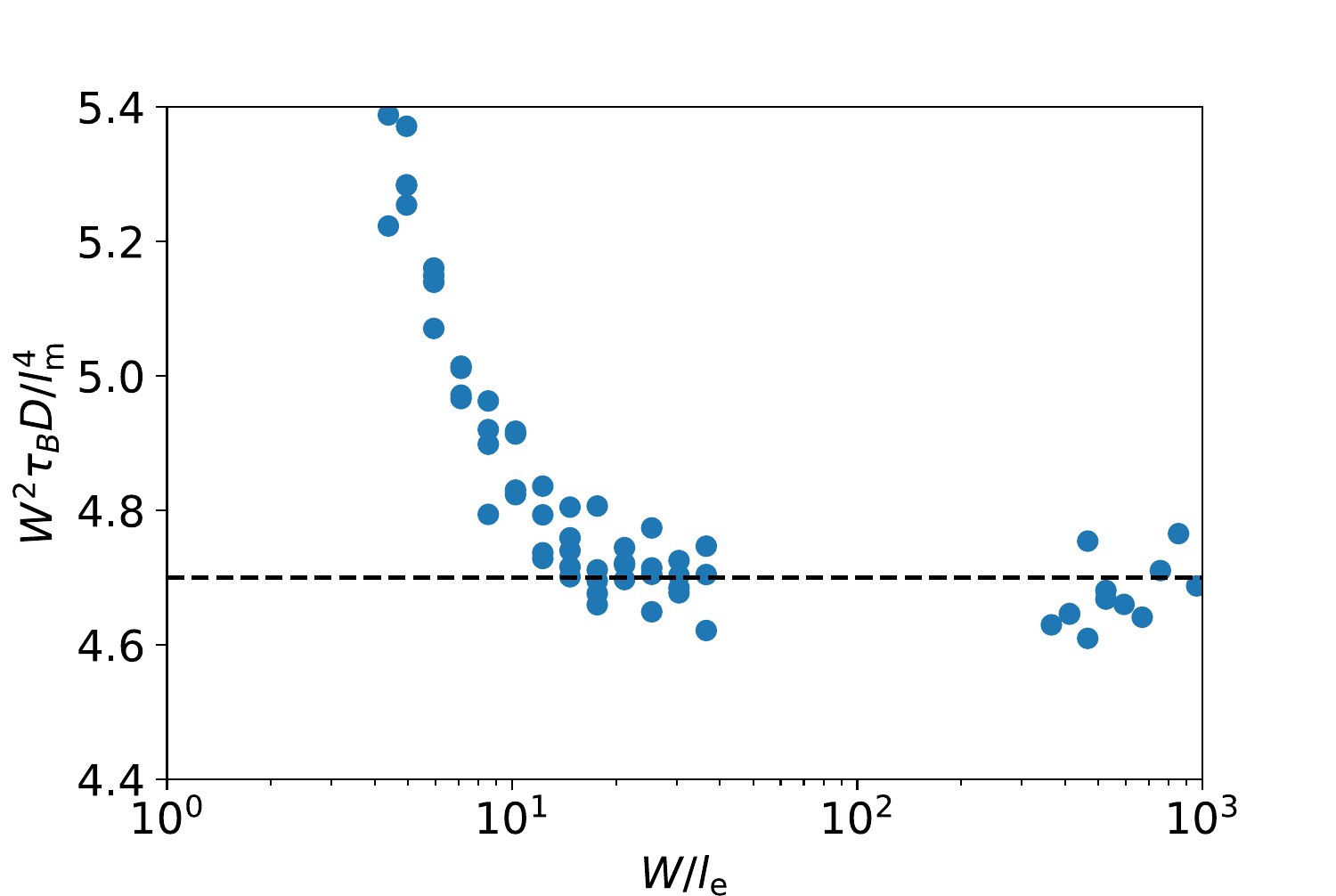}
	\caption{\textbf{Simulating the pre-factor $C$.} Numerical simulation of the magnetic dephasing length $\tau_B$ for a set of parameters $W/l_\text{e}$ and $l_\text{m}/l_\text{e}$. The different data points collapse on a single curve. For $W\gtrsim 10 \cdot l_\text{e}$ we find $C\approx 4.7$, indicated by the dashed line.}
	\label{sim}
\end{figure}

The parametric dependence of the magnetic dephasing length $\tau_B$, used in Eq. 1 of the main text, can be computed numerically by numerically averaging over random electron paths, as explained in detail in Ref.[\citen{van2015spin}]. We apply this method to our hexagonal nanowires of width, $W$, in the diffusive regime, with the mean free path $l_\text{e}$ $<$ $W$. In the diffusive regime, we can expect from general arguments from Ref. [\citen{beenakker1991quantum}] that $\tau_B \sim l_\text{m}^4/D W^2$. In Fig. \ref{sim} we show the results from our numerical simulation, exhibiting indeed this parametric dependence for $W \gtrsim 10 l_\text{e}$. We can then read off the proportionality constant $C$ $\approx$ 4.7. We note that this microscopically computed prefactor differs by a factor of order 1 from $C$ = 3 that would be appropriate for a square cross-section of width $W$[\citen{van2015spin, beenakker1991quantum}]. Since we do not know all microscopic details of our system (such as whether electrons are uniformly occupying the nanowire as assumed in the calculations seen in Fig. \ref{sim}, or are accumulated at the surface), we resort to this previously published value for simplicity.

\section*{S8 - Extracting the effective $\mathbf{g}$-factor\\}

We find the $g$-factor of the hybrid structure from the Zeeman relation, 

\begin{align}
E_{\text{Z}} = g\mu_{\text{B}} B
\end{align}

where $\mu_B$ is the Bohr magneton, $g$ is the Land\'e $g$-factor and B is the magnetic field magnitude. This allows for the factor for this device on the order of $\sim$ 17.

\begin{figure}[htb!]
\centering
	\includegraphics[scale=1]{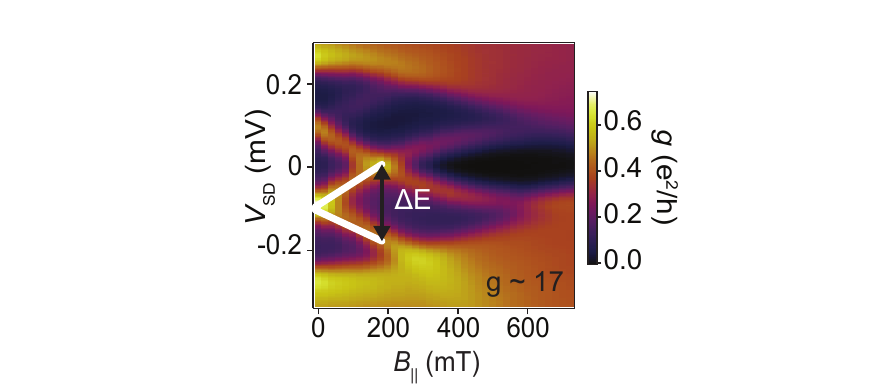}
	\caption{\textbf{Extraction of effective $\mathbf{g}$-factors.} Differential conductance, $g$, as a function of source-drain bias, $V$\textsubscript{SD}, and parallel magnetic field $B_{\parallel}$. The white sloped lines indicate the magnitude of the effective $g$-factor.}
	\label{extractg}
\end{figure}

\bibliography{ref}